\documentclass[submission,copyright,creativecommons]{eptcs}
\providecommand{\event}{WS-FMDS 2012} 
\usepackage{breakurl}             
\usepackage{cite}
\usepackage[cmex10]{amsmath}
\usepackage{algorithmic}
\usepackage{array}
\usepackage{url}
 \usepackage[pdftex]{graphicx}

\title{Using Constraints for Equivalent Mutant Detection\thanks{The research herein is partially conducted within the competence network Softnet Austria II (www.soft-net.at, COMET K-Projekt) and funded by the Austrian Federal Ministry of Economy, Family and Youth (bmwfj), the province of Styria, the Steirische Wirtschaftsf\"orderungsgesellschaft mbH. (SFG), and the city of Vienna in terms of the center for innovation and technology (ZIT).}}
\author{Simona Nica \qquad\qquad Franz Wotawa
\institute{Institute for Software Technology \\
Technische Universit{\"a}t Graz \\ Graz, Austria}
\email{\{snica,wotawa\}@ist.tugraz.at}
}
\def\titlerunning{Equivalent Mutant Detection}
\def\authorrunning{S. Nica \& F. Wotawa}

\begin{document}

\newtheorem{theorem}{Theorem}
\newtheorem{lemma}[theorem]{Lemma}
\newtheorem{corollary}[theorem]{Corollary}
\newtheorem{definition}{Definition}
\newtheorem{claim}{Claim}

\newcommand{\Forall}{\mbox{$\forall\,$}}
\newcommand{\Exists}{\mbox{$\exists\,$}}
\renewcommand{\And}{\mbox{$\,\wedge\,$}}
\newcommand{\Or}{\mbox{$\,\vee\,$}}
\newcommand{\impl}{\mbox{$\,\supset\,$}}
\newcommand{\union}{\mbox{$\,\cup\,$}}


\newenvironment{programlist}%
{\vspace{-0.3em} \begin{tabbing} \quad 12 \=\quad \=\quad \=\quad \=\quad \=\quad \= \kill}%
{\end{tabbing}}

\newcommand{\jade}{\texttt{jade}}
\newcommand{\keyw}[1]{{\textbf {#1}}}
\newcommand{\method}[1]{{\textit {#1}}}
\newcommand{\class}[1]{{\textit {#1}}}
\newcommand{\field}[1]{{\textit {#1}}}
\newcommand{\var}[1]{{\textit {#1}}}
\newcommand{\const}[1]{{\textit {#1}}}
\newcommand{\loc}[1]{{\textit {#1}}}
\newcommand{\category}[1]{{\textit {#1}}}
\newcommand{\protocol}[1]{{\textit {#1}}}
\newcommand{\namespace}[1]{{\textbf {#1}}}
\newcommand{\menu}[1]{{\textit {#1}}}
\newcommand{\file}[1]{{\textit {#1}}}

\newcommand{\stmt}{\texttt}  
\newcommand{\mstmt}{\mathtt}  
\newcommand{\expr}[1]{$\mstmt{#1_{expr}}$}  
\newcommand{\mexpr}[1]{\mstmt{#1_{expr}}}  
\newcommand{\prog}{\Pi}  
\newcommand{\ssa}{\Sigma(\prog)}  
\newcommand{\ssanophi}{\Sigma_{\overline{\Phi}}(\prog)}  
\newcommand{\csp}{CSP(\prog)}  
\newcommand{\maptossa}{\sigma}
\newcommand{\mapfromssa}{\overline{\sigma}}

\newcommand{\annot}{\Arrowvert}
\newcommand{\ass}{\mathcal{A}}

\newcommand{\diag}{\Delta}
\newcommand{\cab}{\Gamma}
\newcommand{\el}{E}
\newcommand{\elim}{\el_{\%}}
\newcommand{\eliminv}{\el_{\%, inv}}
\newcommand{\defeq}{\stackrel{\text{\tiny{def}}}{=}}

\newcommand{\N}{\mathbb{N}}
\newcommand{\R}{\mathbb{R}}
\newcommand{\C}{\mathbb{C}}
\def\Nat{\rm I\hspace*{-0.2em}N}
\newcommand{\J}{\mathbf J}

\newcommand{\algorithmEQMut}{\textbf{equalMutantDetection}}
\newcommand{\algorithmConvert}{\textbf{convert}}
\newcommand{\tool}{\textbf{EqMutDetect}}

\maketitle

\begin{abstract}
In mutation testing the question whether a mutant is equivalent to its program is important in order to compute the correct mutation score. Unfortunately, answering this question is not always possible and can hardly be obtained just by having a look at the program's structure. In this paper we introduce a method for solving the equivalent mutant problem using a constraint representation of the program and its mutant. In particularly the approach is based on distinguishing test cases, i.e., test inputs that force the program and its mutant to behave in a different way. Beside the foundations of the approach, in this paper we also present the algorithms and first empirical results.
\end{abstract}

\section{Introduction}

Mutation testing is a technique used to assess the quality of test suites and is more efficient than other approaches like coverage based metrics \cite{AndrewsBL05}. It is a fault based technique that uses a well defined set of faults. The considered faults can be introduced as slight changes of the original program that lead to a variant of the program which is called mutant. The basic idea behind mutation testing is now to check whether the available test suite is able to detect the mutant. For this purpose the available test suite is executed on these mutated versions of the original program. If there is at least one test run that fails, we say that the mutant is detected or killed. Otherwise, the mutant is alive. A test suite is more efficient than another test suite if it is able to detect more mutants. We measure the efficiency using the \emph{mutation score}. The mutation score is defined as the ratio between the number of mutants detected and the total number of mutants minus the equivalent ones (a mutant is said to be equivalent if syntacticly differs from the original program, but semantically the mutation can not be detected). The ideal mutation score  is 1, i.e., all mutants are successfully detected. For more information on mutation testing we refer the interested reader to \cite{JiaH10}.

One major problem of mutation testing is the {\em equivalent mutant problem}. A mutant is said to be equivalent if there is not such a test case, able to differentiate between the output of the mutant and the output of the original program. When considering the definition of the mutation score, we see that detecting all equivalent mutants is very important. In literature several techniques for equivalent mutant detection exist, e.g., \cite{oc, zeller}. In this paper we also focus on the equivalent mutant problem and describe the underlying techniques of a tool, \tool\ ,  serving our purpose. The tool uses constraint solving for proving the equivalence.

Other approaches are included in \cite{Hierons99usingprogram}, where the authors use program slicing for solving the equivalent mutant problem. In order to show whether a mutant is equivalent or not, the described technique compares the effect of a program and its mutant on certain variables. 
\cite{offpanConstr} introduces a solution most closely to ours, which is also based on constraints. The authors developed a tool that implements a mathematical constraint algorithm. For determining the equivalent mutants, the authors use constraints to recognize infeasible constraints that produce equivalent mutants. The used constraint will describe the circumstances under which a mutant must be detected, i.e. if a test case can detect the mutant, then the constraint system will be true. Otherwise, no test case is able to kill the mutants, and, therefore, an equivalent mutant is detected. The authors state that a mutation is detected only when it satisfies three conditions: reachability, necessity and sufficiency. Unlike in our approach, where we make use of the distinguishing test case concept, the authors propose three strategies to detect equivalent mutants: negation, constraint splitting and constants comparison. For negation, the authors use constraint negation and partial negation of two constraints, e.g., C1 and C2, rewrite one of the two constraints and then compare them. If they are syntactically equal, then the constraint system is infeasible and a mutant with this infeasible constraint system is equivalent. But constraint negation does not help in determining if the necessity constraint conflicts with the path expression. Therefore, in order to detect conflicts, constraint splitting is introduced. Constant comparison is based on a property common in constraints generated for test cases: both constraints should have the format \textit{(V RelOp K)}, where \textit{V} is a variable, \textit{RelOp} is a relational operator, and \textit{K} is a constant. Also the variables in both constraints
are required to be the same.
   
\cite{zeller} uses the impact on executions and also on the return values. The approach examines the impact of mutations at the coverage level. In particular the approach makes use of observations about the lines of code executed. The coverage from the original execution is  compared with the coverage of the mutants. Regarding the impact over the return values, the authors established two types of non-equivalent mutations, which do not affect the coverage level: mutations which affect only the data, but not the output values of a program, and mutations reflected in the return values.

In this paper we introduce an approach that makes use of the constraint representations of a program and its mutant. The idea is to use the constraints for computing distinguishing test cases, i.e., test cases that allow for differentiating two program runs using the same input. We also discuss challenges and give some initial results when applying our approach to smaller programs. 

The paper is organized as follows. In Section II we present the basic definitions followed by an  introduction of the program's constraint representation in Section III. We describe the underlying algorithms in Section IV. In Section V we give an overview of our tool, and present the first empirical results. In Section VI we conclude the paper.

\section{Basic definitions}

\begin{figure}[t]
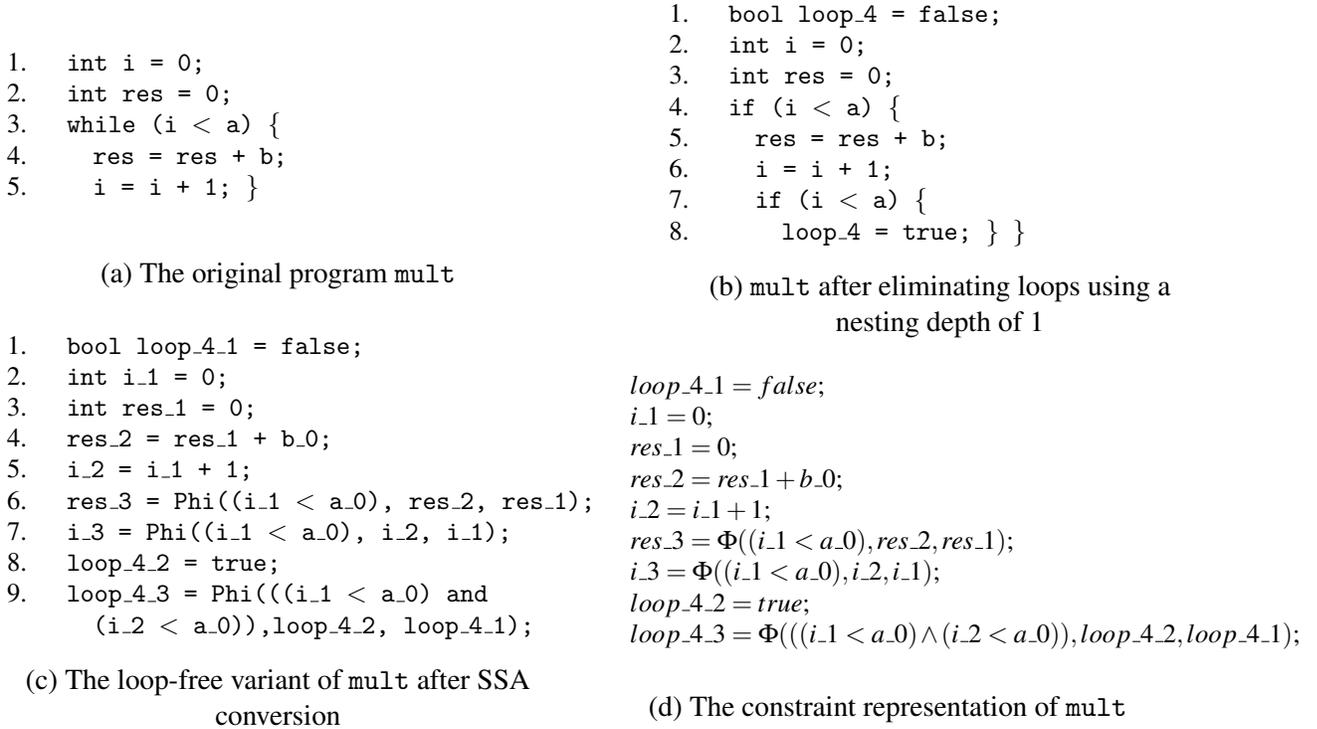


\begin{minipage}{0.45\linewidth}
\centering
{\small 
\begin{programlist}
1. \> \tt int i = 0; \\
2. \> \tt int res = 0; \\
3. \> \tt while (i $<$ a) \{ \\
4. \> \> \tt res = res + b; \\
5. \> \> \tt i = i + 1; \}  \\
\end{programlist}
}

(a) The original program {\tt mult} \\[1em]
\end{minipage}
~
\begin{minipage}{0.45\linewidth}
\centering
{\small 
\begin{programlist}
1. \> \tt bool loop\_4 = false; \\
2. \> \tt int i = 0; \\
3. \> \tt int res = 0; \\
4. \> \tt if (i $<$ a) \{ \\
5. \> \> \tt res = res + b; \\
6. \> \> \tt i = i + 1; \\
7. \> \> \tt if (i $<$ a) \{ \\
8. \> \> \> \tt loop\_4 = true; \} \}  
\end{programlist}
}

(b) {\tt mult} after eliminating loops using a nesting depth of 1\\[1em]
\end{minipage}
~
\begin{minipage}{0.45\linewidth}
\centering
{\small 
\begin{programlist}
1. \> \tt bool loop\_4\_1 = false; \\
2. \> \tt int i\_1 = 0; \\
3. \> \tt int res\_1 = 0; \\
4. \> \tt res\_2 = res\_1 + b\_0; \\
5. \> \tt i\_2 = i\_1 + 1; \\
6. \> \tt res\_3 = Phi((i\_1 $<$ a\_0), res\_2, res\_1); \\
7. \> \tt i\_3 = Phi((i\_1 $<$ a\_0), i\_2, i\_1); \\
8. \> \tt loop\_4\_2 = true;  \\
9.\> \tt loop\_4\_3 = Phi(((i\_1 $<$ a\_0) and \\
\> \> \tt (i\_2 $<$ a\_0)),loop\_4\_2, loop\_4\_1); 
\end{programlist}
}

(c) The loop-free variant of {\tt mult} after SSA conversion \\[1em]
\end{minipage}
~
\begin{minipage}{0.45\linewidth}
\centering
{\small 
$$\begin{array}{l}
loop\_4\_1 = false; \\
i\_1 = 0; \\
res\_1 = 0; \\
res\_2 = res\_1 + b\_0; \\
i\_2 = i\_1 + 1; \\
res\_3 = \Phi((i\_1 < a\_0), res\_2, res\_1); \\
i\_3 = \Phi((i\_1 < a\_0), i\_2, i\_1); \\
loop\_4\_2 = true;  \\
loop\_4\_3 = \Phi(((i\_1 <  a\_0) \And (i\_2 < a\_0)),loop\_4\_2, loop\_4\_1); 
\end{array}$$
}

(d) The constraint representation of {\tt mult}
\end{minipage}

\caption{A program {\tt mult} for computing $a * b$ and its conversion to a set of constraint. Note that $a$ and $b$ are positive integers including zero and are used as input variables, and variable $res$ denotes the result.}
\label{fig:conv_ex}
\end{figure}

We start by stating the basic definitions we will use along this paper. For this purpose we explain mutants and state the equivalent mutant problem. For illustration purposes we make use of the small program that is depicted in Figure~\ref{fig:conv_ex}~(a). The program implements the multiplication of two natural numbers. We do not define the underlying programming language formally. We assume a sequential programming language without object-oriented constructs. Also, in our first algorithm implementation, we do not consider procedure calls, neither recursion.  However, the general idea of detecting equivalent mutants can be applied to other languages requiring specialized constraint solver.

We start with the definition of a test case that specifies the expected output behavior given an input. In sequential languages the input and the output can be specified as variable environments. A variable environment is a mapping from a variable to its value. Note that in the following definition it is not necessary that the values of all output variables are given.

\begin{definition}\textbf{[Test Case]}
We define a test case for a program $\Pi$ as a set $(I,O)$ where $I$ is the input variable environment specifying the values of all input variables used in $\Pi$, and $O$ the output variable environment. If no output variable environment is specified, we set $O$ to $\emptyset$. 
\end{definition}

A \textit{failing test case} is a test case for which the output environment computed from the program $P$, executed over input $I$, is not consistent with the expected output. Otherwise, the test case is said to be a \textit{passing test case}. The test case $(\{({\tt a},1),({\tt b},2)\}, \{({\tt res},2)\})$ is a passing test case for the multiplication program from Fig.~\ref{fig:conv_ex}~(a).

A \textit{test suite $TS$} for a program $\Pi$ is a set of test cases for $\Pi$. 

Moreover, we introduce the concept of distinguishing test cases.

\begin{definition}\textbf{[Distinguishing Test Case]}
Given programs $\Pi_1$ and $\Pi_2$. We say that a test case $(I,O)$ is a distinguishing test case, if the output variable environments computed by $\Pi_1$ and $\Pi_2$ using the same input $I$ are different.
\end{definition}

We now briefly introduce mutants and state the equivalent mutant problem.

\begin{definition}\textbf{[Mutant]}
Given a program $\Pi$ and a statement $S_\Pi \in \Pi$. Let $S'_\Pi$ be a statement that results from $S_\Pi$ when applying changes like modifying an operator or a variable. Then program $M$ is the mutant of program $\Pi$ with respect to statement $S_\Pi$, obtained when replacing $S_\Pi$ with $S'_\Pi$.
\end{definition}

A mutant of program {\tt mult} (Fig.~\ref{fig:conv_ex}~(a)) would be a program where Line 5 is changed to {\tt i = i + 2;}. 

Other mutants can also be generated by changing constants, variables, or other operators in the program. From here on we assume that there exists a function that generates mutants by applying small changes to the original program. Such a function delivers a set of mutants that are syntactically different to the original program. Unfortunately, syntactical difference does not guarantee that the mutant behaves differently to the original program. 

\begin{definition}\textbf{[Equivalent Mutant]}
Given a program $\Pi$, and one of its mutants $M$. We say that $M$ is an equivalent mutant if semantically $M$ behaves exactly like $\Pi$. If we consider the \textit{distinguishing test case} definition, for the equivalent mutant we will not detect a test case, able to point out the difference between the original program $\Pi$ and its corresponding mutant, $M$.   
\end{definition}

The {\em equivalent mutant problem} is a decision problem that allows to determine whether a program is behavioral equivalent to its mutant. This problem is obviously equivalent to the program equivalence problem that is well known to be undecidable in general.

\section{Representing programs as constraints}

In order to be self-contained we briefly recall the conversion of sequential programs into their equivalent constraint representation under certain assumptions. For the conversion of programs to their constraint representation we make use of previous work, conducted in the constraint based testing research area \cite{GotliebATD, Collavizza, Denmat07improvingconstraint, euclide}. We refer the interested reader to \cite{wot2010b} for a more detailed introduction of the conversion where the constraint representation is used for fault localization. We take again the small program from Figure~\ref{fig:conv_ex}~(a) as an example to demonstrate the conversion.  

The first step of the conversion process is to eliminate all loop statements. The idea here is to replace a while statement with a nested conditional statement of a pre-defined nesting depth $nd$.  Obviously the value of $nd$ determines whether the original program behaves equivalent to its corresponding loop-free variant. If we choose a value for $nd$ that is too small for a given test case, then the programs behave in a different way. On the other hand when choosing a very large value for $nd$ the resulting loop-free program becomes unnecessarily large. Therefore in practice a trade-off for $nd$ is necessary, together with means for detecting situations where $nd$ is too small. We ensure this by introducing a fresh boolean variable {\tt loop}$_i$ for each loop, that is initialized with {\tt false} and set to {\tt true} whenever nesting depth $nd$ is not large enough. 

In the second step the loop-free program is converted to its static single assignment form (SSA) \cite{brandis94}. In the SSA every variable is defined only once. This can be ensured by mapping each variable {\tt x} occurring in a program to a variable {\tt x\_}$i$ where $i$ represents an index variable starting from 0. Every time {\tt x} is defined, the index $i$ is increased by 1 and added to the variable. If the variable is referenced after the current index, $i$ is used until a new re-definition of {\tt x} occurs. Beside adding indices there is one additional conversion rule for conditional statements. Let us assume the following program fragment: {\small \tt if (x $>$ 4) \{ y = 0; \} else \{ y = 1;  \} }. After adding indices to the variable let us assume the following situation:
{\small \tt if (x\_0 $>$ 4) \{ y\_1 = 0; \} else \{ y\_2 = 1;  \} }. In this case whether to use {\tt y\_1} or {\tt y\_2} after the conditional, when referring to {\tt y}, is not defined. In order to overcome this problem, conditional statements are replaced with a $\Phi$ function. The purpose of the $\Phi$ function is to map the variables defined in either of the branches of a conditional statement to the same variable but with a new index that can be referenced after the statement. For our program fragment this additional rule leads to the following program:

{\small
\begin{programlist}
\tt y\_1 = 0; \\
\tt y\_2 = 1; \\
\tt y\_3 = Phi((x\_0 $>$ 4),y\_1,y\_2); 
\end{programlist}
}

Note that we assume a function {\tt Phi} available in the programming language representing the function $\Phi$, which is defined as follows: $\Phi(C,x_1,x_2) = \left\{
\begin{array}{ll} 
x_1 & \mbox{if~$C$ evaluates to {\tt true}} \\
x_2 & \mbox{otherwise}\\
\end{array}\right.$

It is also worth noting that after the second step, the program comprises assignment statements only. Therefore in the final step of conversion we only need to map the assignment statements to equations (or constraints). This step has to take care of the underlying constraint solver. In our case we use MINION \cite{min}. Because of space restrictions we do not discuss the mapping to MINION programs. Instead we use mathematical equations written in italics for representing the arrays, and we assume a constraint solver that is able to compute a solution for a set of equations. From here on we also assume that a function {\bf convert($\Pi$,$nd$)} implements the conversion taking a program $\Pi$ and a maximum nesting depth $nd$ as inputs. Figure~\ref{fig:conv_ex} (b)-(d) depicts the conversion results step-by-step when applying {\bf convert} to the multiplication program from Figure~\ref{fig:conv_ex}~(a) and using $nd=1$ as maximum nesting depth.

\section{The equivalent mutant detection algorithm}

The basic idea behind the equivalent mutant detection algorithm is motivated by the concept of distinguishing test cases introduced for program debugging in \cite{wot2010b}. In this paper the authors describe how distinguishing test cases can help to further improve fault localization via extending the available test suite. The equivalent mutant problem can be solved as a byproduct of computing distinguishing tests. If there is no test case that distinguishes the program from its mutant, then the program and its mutant have to be equivalent. Otherwise, it is not possible to state that both programs must be different with respect to their behavior. 

This is due to the fact that the constraint representation of a program, as introduced in the previous section, is only behavioral equivalent to the corresponding program, for test cases not exceeding the number of considered iterations in while statements. This fact is less problematic in pure fault localization where no further test cases are generated, because the number of iterations to be considered for generating the loop-free program is known in advance. This holds because fault localization requires at least one failing test case, which determines the number of iterations assuming that the program halts. 

As a consequence of this observation it is not always possible to determine equivalent mutants when using distinguishing test cases. However, this is not a surprise because of the undecidability of the equivalent mutant problem. So what are the consequences? In case a distinguishing test case can be computed, we have to check whether this test case is a feasible test case or not. Feasible means that the program and its mutant are able to execute the input of the test case and return the expected output, which can be derived from the solution of the corresponding constraint representation. If the obtained distinguishing test case is feasible, the programs are not equivalent. Otherwise, we have to search for a different distinguishing test case. We can do this by adding the information that the previously computed input is not allowed to be computed anymore. Moreover, there is a second problem due to the number of iterations considered during conversion. If the constraint solver returns no solution, this might be due to the chosen nesting depth. Therefore, we have to increase the nesting depth and start searching again. Of course this cannot be done forever. Therefore in practice we limit the nesting depth to a pre-defined maximum value.

The following algorithm implements the underlying basic idea of checking whether a mutant is equivalent to its corresponding program or not. 

~\\[-1em]
{\bf Algorithm equalMutantDetection($\Pi$,$M$,$nd$, $nd_{max}$}) \\
{\em Input:} A program $\Pi$, its mutant $M$, the initial nesting depth $nd$, and the maximum nesting depth $nd_{max}$.\\
{\em Output:} {\tt true} if $\Pi$ is equivalent to $M$ and {\tt false}, otherwise. 
\begin{enumerate}
\item \label{label1} Convert the program into its constraint representation: $CON_\Pi = \mbox{\bf convert}(\Pi,nd)$
\item Let $M'$ be a program obtained from $M$ by adding the postfix {\tt \_M} to all variables. 
\item Convert the mutant into a set of constraints:Ê$CON_M = \mbox{\bf convert}(M',nd)$
\item Let $CON$ be $CON_\Pi \cup CON_M$.
\item For all input variables {\tt x} of $\Pi$, add the constraint $x = x\_M$ to $CON$.
\item Let {\tt y$^1$}, \ldots,  {\tt y$^k$} be the $k$ output variables of $\Pi$. Add the constraint $y^1 \not= y^1\_M \Or \ldots \Or y^k \not= y^k\_M$ to $CON$.
\item \label{loop2} Call a constraint solver on $CON$ and let $SOL$ be the set of solutions, e.g., mappings of variables to values that satisfy all constraints in $CON$.
\item If there exists no solution $SOL$, i.e., $SOL = \emptyset$, then $\Pi$ and $M$ are potentially equivalent. In this case do the following:
\begin{enumerate}
\item If $nd \geq nd_{max}$, terminate the algorithm and return {\tt true}.
\item Otherwise, increase $nd$ by 1 and go to \ref{label1}.
\end{enumerate}
\item Otherwise, there exists a (non-empty) solution $SOL$. If there is no variable $loop_j$ with an assigned value of {\tt true} in $SOL$, then return {\tt false}. Otherwise, add the information that the inputs computed in $SOL$ are not valid to $CON$ and go to \ref{loop2}.
\end{enumerate}

In the {\bf equalMutantDetection} algorithm lines 1--6 are for constructing the constraint system. In Line 5 constraints are added in order to force the input variables of the program and its mutant to be the same. Line 6 is for stating that a distinguishing test case triggers at least one output variable to hold a different value after execution. Note also that the mutant variables are changed before conversion. 
Line 7 calls the constraint solver. In Line 8 it is handled the case where no distinguishing test case can be found. This either leads to the result that the mutants are equivalent or to an increase of the considered nesting depth and a re-computation. Line 9 is for handling the case where a distinguishing test case is found. There feasibility is checked via testing whether the number of iterations for computing a solution has been exceeded. If the computed test case is not feasible, search starts again using the information already gathered. 

Note that the algorithm always terminates because of the introduced $nd_{max}$ variable. However, it can be the case that the algorithm returns {\tt true} but there is a distinguishing test case requiring the programs to be executed for a longer time. However, if the algorithm returns {\tt false}, the program and its mutant are definitely different with respect to their semantics. This restriction comes from the undecidability of the underlying equivalent mutant problem.

\begin{table}[t]
\centering
\scriptsize
\begin{tabular}{|c||c|c|c|} 
\hline 
\textbf {\bf Class } & \bf{LOC} & $\bf{No_{Mut}}$ & $\bf{Det_{EqMut}}$  \\ 
\hline
tcas01&125 & 231& 137\\  \hline
tcas02&125 & 231& 139\\  \hline
tcas03&125 & 231& 149\\  \hline
tcas04&125 & 231& 142\\  \hline
\ldots & \ldots & \ldots & \ldots \\ \hline
tcas25&125 & -& -\\  \hline
\ldots & \ldots & \ldots & \ldots \\ \hline
tcas41&125 & 229& 119\\  \hline\hline
\bf AVG    & \bf 125 & \bf 226  & \bf 137 \\ \hline
\end{tabular}
\begin{tabular}{|c||c|c|c|} 
\hline 
\textbf {\bf Class } & \bf{LOC} & $\bf{No_{Mut}}$ & $\bf{Det_{EqMut}}$  \\ 
\hline
C432Order &382 & 2038 & 1341 \\  \hline
ArrayOperations & 101& 496& 11 \\  \hline
BubbleSort&18& 62 & 9\\  \hline
CalculateRectArea&14 &18 & 2 \\  \hline
CalculateRectPerimeter &13 & 21 & 4\\  \hline
CoffeeMachine & 43& 91 & 0 \\  \hline
FindEvenOrOddNumber&15& 34&  3\\  \hline
FindLargestSmallestNumber&19  &41 & 13 \\  \hline
GcdATC&35 & 95& 29\\  \hline
NumberFactorial &19 & 38& 12\\  \hline
\end{tabular}
\caption{EqMut Detection}
\label{EqDetectResults}
\end{table}

\section{Empirical results}

In order to verify the practicability of our approach we have partially implemented the \algorithmEQMut\ algorithm, ignoring increasing the bound of the nesting depth. This extension will be implemented in the next version of our tool \tool\, which makes use of the MuJava \cite{SeungMa} mutant generation tool and the MINION \cite{min} constraint solver. We aim to offer a reliable mutation test case generation tool that handles the equivalent mutant problem as good as possible. We developed  \tool\ in Java. We also modified MuJava in order to be JDK 1.5 compliant (see \cite{NP}). \tool\ comprises three main components: (1) Mutant generation using MuJava, (2) Equivalent mutants detection and reduction, and (3) Test case generation.
The currently used mutation tool produces method level and class level mutations. However, in our research experiments, we excluded object oriented programs and generate only method level mutations, i.e., we make use of the method-level operators defined for MuJava, and mutate statements by replacing, deleting and inserting primitive operators, e.g., arithmetic operator, relational operator, etc.

Using the implemented tool \tool\ we conducted first experiments. For this purpose we applied \tool\ on some smaller programs varying from 13 to more than 380 lines of code. The used programs include the TCAS files \cite{tcas}, and programs that implement operations on arrays, and  mathematical operations. The TCAS files were converted to Java syntax, in order to produce mutations with MuJava. In our experiments the nesting depth for representing loops vary from 2 to 5. For constraint solving we have established a 5 minutes time bound. We executed \tool\ using the maximum and minimum nesting depth, and observed no difference regarding the obtained solutions. Therefore, we conclude that for the current experiments a small number of iterations is adequate.

In Table~\ref{EqDetectResults} we summarize the obtained results. {\bf LOC} denotes the lines of code, $\bf{No_{Mut}}$ represents the total number of computed mutants, and $\bf{Det_{EqMut}}$ is the number of equivalent mutants detected using \tool. We observe from Table~\ref{EqDetectResults} that in some cases no equivalent mutants were generated, i.e., the constraint solver was always able to find a distinguishing test case. We  also encountered the situation where the constraint solver tried to find a solution within the predefined time limit, but it did not manage. In this case we terminated search.   
  
Regarding the quantitative results of our experiments, there are some interesting findings. About 40\% of the mutants generated for the \textit{tcas} files were found equivalent when using our approach. This happens because we only take into account mutations, which are reachable and have an impact on the return values. Also concerning the tcas files, we found situations when we could not apply our approach, because the mutation engine failed to generate mutants, e.g., for \textit{tcas25}. It is also interesting to note that for the largest program {\em C432Order} the constraint solver was able to provide solutions within the 5 minutes bound.

\section{Conclusion}

Reducing the number of equivalent mutants plays a significant role in determining the efficiency of test suites. The mutation score can be significantly improved when removing all equivalent mutants. Hence, in this paper we discussed an approach for detecting equivalent mutants.
In the presented approach we combine constraint representations of programs with mutation testing. In particular we are searching for a distinguishing test case in order to differentiate the program from its mutant. Because of the fact that the underlying problem is undecidable the approach does not guarantee to find a solution. The influence of certain parameters like the given nesting depth is left for future research. As a byproduct the approach allows for adding new test cases to the test suite. Computed distinguishing test cases can be used to increase the mutation score.

In our current experiments we consider smaller programs. Therefore, the efficiency and scalability of the approach can be disputed. Hence, future research will include improving the experimental bases and using more and larger programs. Moreover, we also want to extend the tool in order to handle object-oriented programs.

\bibliographystyle{eptcs}
\bibliography{eptcsbibliografy}

\end{document}